\documentclass[fleqn,12pt]{wlscirep}

\usepackage[utf8]{inputenc}
\usepackage[T1]{fontenc}
\usepackage{chemformula}

\usepackage{adjustbox}
\usepackage{lipsum}
\usepackage{mathtools}
\usepackage{physics}
\usepackage{floatrow}
\usepackage{hyperref}
\usepackage[justification=centering]{caption}
\usepackage{amsmath}
\usepackage{nccmath}
\usepackage{subcaption}
\usepackage{floatrow}
\usepackage[thinlines]{easytable}
\usepackage{cleveref}
\usepackage{gensymb}
\usepackage{booktabs}  
\usepackage{ltablex}
\usepackage{bbm}
\DeclareMathOperator{\sgn}{sgn} 
 
\usepackage{graphicx} 
\usepackage{subcaption} 
 
\usepackage{soul}
\usepackage{cancel}

\newcommand\sbullet[1][.5]{\mathbin{\vcenter{\hbox{\scalebox{#1}{$\bullet$}}}}}  
 
\title{Hypomagnetic field effects as a potential avenue for testing the radical pair mechanism in biology }

\author[1,2,3,*]{Hadi Zadeh-Haghighi}
\author[1,2,3]{Rishabh Rishabh}
\author[1,2,3,*]{Christoph Simon}

\affil[1]{Department of Physics and Astronomy, University of Calgary, Calgary, AB, T2N 1N4, Canada}
\affil[2]{Institute for Quantum Science and Technology, University of Calgary, Calgary, AB, T2N 1N4, Canada}
\affil[3]{Hotchkiss Brain Institute, University of Calgary, Calgary, AB, T2N 1N4, Canada}

\affil[
*]
{hadi.zadehhaghighi@ucalgary.ca \& csimo@ucalgary.ca}

\begin{abstract}

Near-zero magnetic fields, called hypomagnetic fields, are known to impact biological phenomena, including developmental processes, the circadian system, neuronal and brain activities, DNA methylation, calcium balance in cells, and many more. However, the exact mechanism underlying such effects is still elusive, as the corresponding energies are far smaller than thermal energies. It is known that chemical reactions involving radical pairs can be magnetic field dependent at very low intensities comparable to or less than the geomagnetic field. Here, we review in detail hypomagnetic field effects from the perspective of the radical pair mechanism, pointing out that under certain conditions, they can be comparable or even stronger than the effects of increasing the magnetic field. We suggest that hypomagnetic field effects are an interesting avenue for testing the radical pair mechanism in biology.
 
\textbf{Keywords:} \textit{hypomagnetic field effects in biology, low field effects, radical pair mechanism, quantum spin, spin chemistry} 

\end{abstract}

\begin{document}

\maketitle

\section{ Introduction}

Weak magnetic field effects are widespread in biology \cite{Zadeh_Haghighi_2022}. The strength of the magnetic field in such phenomena can go below the magnitude of the geomagnetic field. Hypomagnetic fields, which are near-zero fields and can be produced by shielding the earth's magnetic field, are also known to impact biological functions \cite{Belyavskaya2004,Maffei2014,Binhi2017,Zhang2020,Zhang2021rev,Xue2021,TeixeiradaSilva2015,Tsetlin2016}, such as plants' growth, development, and evolution \cite{Belyavskaya2004,Maffei2014,TeixeiradaSilva2015,Tsetlin2016,Parmagnani_2022}, reactive oxygen species \cite{Zhang2020}, circadian rhythm \cite{Xue2021}, learning and memory \cite{Zhang2021b} and so on. Such effects, which are also often referred to as low field effects (LFEs), have been observed in genetics, developmental processes, the circadian clock, neurons and the brain, and so forth. It has also been suggested that extinction events on Earth \cite{Raup1984} may be related to the changes in geomagnetic field \cite{LIPOWSKI2006}. Early studies, motivated by the concerns around the health of astronauts in outer space, concluded that exposure to hypomagnetic fields had adverse effects on human health \cite{becker1963relationship,Beischer1971,beischer1967exposure,Dubrov1978}. Apart from hypomagnetic field effects on animal and human cells and tissues, deprivation in the geomagnetic field can influence the development of plants as well \cite{TeixeiradaSilva2015,Tsetlin2016}. It, therefore, seems pertinent to conclude that the geomagnetic field may play essential roles in living organisms, and diminishing it or making it disappear could result in adverse consequences.

\par
Earth's magnetic field, depending on latitude, ranges from $\sim$24 to $\sim$66 $\mu$T \cite{Alken2021}, which is hundreds of times smaller than the strength of a typical refrigerator magnet. Remarkably, the scale of energies corresponding to the geomagnetic field is about a million times smaller than thermal energies, $k_B T$ ($k_B$ and $T$ are the Boltzmann constant and temperature, respectively), at biologically relevant temperatures. This makes any classical interpretation of hypomagnetic field effects in biology highly challenging. Thus far, the mechanisms behind such effects remain elusive \cite{Nordmann_2017,Kirschvink_2010}. It must also be noted that classical models based on stochastic resonance \cite{Galvanovskis_1997}, ion channels \cite{Rosenspire_2005}, and magnetic induction \cite{Nimpf_2019} have also been proposed for weak magnetic field effects in biology. However, for instance, in Ref \cite{Hochstoeger_2020}, the authors show that the experimental observation supports the radical pair mechanism based on cryptochrome for avian magnetoreception.
\par
It is known that applied magnetic fields can influence reaction rates and yields in certain chemical reactions 
\cite{Fessenden_1963,Sagdeev_1973,Brocklehurst_1974,Timmel1998}. The key elements in these processes are pairs of radicals--transient molecules with an odd number of electrons in their outer shell--that carry quantum spins. Electrons possess spins of $S=\dfrac{1}{2}$. Spin also has a magnetic feature, meaning that any other spins and magnetic field around can influence the particle's spin state (radical). A pair of radicals can be formed by electron transfer between two molecules or by breaking a chemical bond. A radical pair can be in a singlet or triplet state, depending on the donor and acceptor molecule’s spin configuration \cite{Hayashi2004}. Spin is usually a conserved quantity in reactions involving organic molecules, which is essential for magnetic field effects in biochemical reactions involving radical pairs \cite{Steiner1989}. In other words, depending on the spin state of the radical pair, the chemical process will take either singlet or triplet paths, resulting in different chemical yields.
\par
It has been proposed that the radical pair mechanism underlies many magnetic field effects across biology \cite{Zadeh_Haghighi_2022}. Radical pair models have been proposed for LFEs \cite{Timmel1998}, including avian magnetoreception \cite{Ritz2000,Xu2021}, isotope effects on xenon anesthesia \cite{Smith2021} and lithium effects on mania \cite{Zadeh2021Li}, and magnetic field effects on the circadian clock \cite{Zadeh2022CC}. In recent studies, it is also suggested that radical pairs may explain hypomagnetic field effects on hippocampal neurogenesis \cite{rishabh2021radical} and microtubule reorganization \cite{ZadehHaghighi2022}. The LFEs on radical pair reactions have been extensively studied experimentally and theoretically \cite{Lewis_2018,Barnes_2014,W_Eveson_C_R_Timmel_B_Brockle_2000,Timmel_2004,Maeda_2008,Hore2019}. Here, based on the radical pair mechanism, we show that changing the magnetic field experienced by a biological system from the geomagnetic field to near zero can, under certain conditions, lead to stronger effects than changing it from the geomagnetic field to a much stronger field (Fig. \ref{fig:SY-B}). We hence suggest that hypomagnetic field effects are an interesting way of testing radical pair models.

\section{Results}
\subsection{Spin dynamics}
Starting from a singlet state, the initial spin density matrix for the radical pair reads as follows:
\begin{ceqn}
\begin{equation}
\hat{\rho}(0) = \dfrac{1}{M} \hat{P}^S, 
\label{eq:rho0S}
\end{equation}
\end{ceqn}
and for triplet initial state:
\begin{ceqn}
\begin{equation}
\hat{\rho}(0) = \dfrac{1}{3M} \hat{P}^T, 
\label{eq:rho0T}
\end{equation}
\end{ceqn}
where
\begin{ceqn}
\begin{equation}
\hat{P}^S= \ket{S} \otimes \bra{S} \otimes \hat{\mathbbm{1}}_{M},
\label{eq:PS}
\end{equation}
\end{ceqn}

\begin{ceqn}
\begin{equation}
\hat{P}^T= \big\{ \ket{T_+} \otimes \bra{T_+} + \ket{T_0} \otimes \bra{T_0} + \ket{T_-} \otimes \bra{T_-}\big\} \otimes \hat{\mathbbm{1}}_{M},
\label{eq:PT}
\end{equation}
\end{ceqn}

\begin{ceqn}
\begin{equation}
\hat{P}^S + \hat{P}^T= \hat{\mathbbm{1}}_{4M},
\label{eq:PS+PT}
\end{equation}
\end{ceqn}

\begin{ceqn}
\begin{equation}
M =\prod_{i}^n (2 I_i +1).
\label{eq:m}
\end{equation}
\end{ceqn}

Here $\hat{P}^S$ and $\hat{P}^T$ are the singlet and triplet projection operators, respectively, $M$ is the nuclear spin multiplicity, $I_i$ is the spin angular momentum of $i$-th nucleus, and $\hat{\mathbbm{1}}_n$ is the $n\times n$ identity matrix.
\par
For studying the spin dynamics of the radical pairs, it is essential to consider the interactions between the unpaired electron spins on each radical with each other, surrounding nuclear spins, and external magnetic fields. All these interactions can be included in the Hamiltonian of the system, $\hat{H}$. The next step is solving the density matrix equation: 
\begin{ceqn}
\begin{align}
\dfrac{d \hat{\rho}(t)}{dt} = -\dfrac{i}{\hbar}[\hat{H},\hat{\rho}(t)],
\label{eq:master}
\end{align}
\end{ceqn}
where $\hat{\rho}(t)$ and $\hat{H}$ are the spin density and Hamiltonian operators, respectively, $i$ is the imaginary unit and $\hbar$ is the reduced Planck constant. [·,·] denotes the commutator. 
\par
For simplicity, let us consider only the isotropic hyperfine ($\hat{H}_{HFI}$) and Zeeman ($\hat{H}_Z$) interactions. We neglect dipolar and exchange interactions; it should be noted that including these terms in the model can reduce the magnetic field sensitivity \cite{Hogben2009,rishabh2021radical}. However, this may be compensated by potential amplification effects in the biological systems \cite{Player2021,Zadeh2022CC}. Of note, the anisotropic components of the hyperfine interactions are only relevant when the radicals are aligned and immobilised\cite{Schulten1978}. The Hamiltonian for this system reads as follows:

\begin{ceqn}
\begin{align}
\hat{H}=\hat{H}_Z +\hat{H}_{HFI} =  - \gamma_e h \big(\mathbf{\hat{S}}_A+\mathbf{\hat{S}}_B\big)\sbullet \mathbf{B}+\sum^{N_A}_{i} a_{iA} \mathbf{\hat{S}}_A \mathbf{\hat{I}}_{iA}+\sum^{N_B}_{i} a_{iB} \mathbf{\hat{S}}_B \sbullet\mathbf{\hat{I}}_{iB},
\label{eq:ham}
\end{align}
\end{ceqn}
where $\gamma_e$, $h$, $\mathbf{\hat{S}}_{A/B}$, $\mathbf{B}$, $a_{iA/B}$, $\mathbf{\hat{I}}_{iA/B}$, and $N_{A/B}$ are the electron magnetogyric ratio, the Planck constant, the spin operators of electron $A/B$, the magnetic field, the isotropic hyperfine coupling constants for electron $A/B$, nuclear spin of $i$-th nucleus for electron $A/B$, and the number of nuclear spins coupled to electron $A/B$, respectively.

Let us consider a minimal model comprising two electron spins, $A$ and $B$, where $A$ is coupled to a spin-1/2 nucleus, $\mathbf{\hat{S}}_{A}\otimes \mathbf{\hat{I}}_{A} \otimes \mathbf{\hat{S}}_{B}$, and the magnetic field has only $z$ component, $B_z$. In this case, the Hamiltonian has the following form:

\begin{gather}
\centering
\hat{H}
 =
  \begin{bmatrix}
\dfrac{a}{4}+B_z & 0& 0& 0& 0& 0& 0& 0 \\
    0&\dfrac{a}{4}&0&0&0&0&0&0 \\
    0&0&-\dfrac{a}{4}+B_z&0&\dfrac{a}{2}&0&0&0 \\
    0 & 0& 0& \dfrac{a}{4}-B_z& 0& \dfrac{a}{2}& 0& 0 \\
    0&0&\dfrac{a}{2}&0&-\dfrac{a}{4}&0&0&0 \\
    0&0&0&\dfrac{a}{2}&0&-\dfrac{a}{4}-B_z&0&0 \\
    0 & 0& 0& 0& 0& 0& \dfrac{a}{4}& 0 \\
    0&0&0&0&0&0&0&\dfrac{a}{4}-B_z   
   \end{bmatrix}
\end{gather}
where $a$ is the hyperfine coupling constant. The normalized eigenstates of this system, $\psi_i$, are the following: 

\begin{ceqn}
\begin{align}
 \label{eq:eigenstates}
 \begin{gathered}
\psi_1=\ket{\uparrow}_A \otimes \ket{\uparrow}_N\otimes \ket{\uparrow}_B,
\\
\psi_2=\ket{\uparrow}_A \otimes \ket{\uparrow}_N\otimes \ket{\downarrow}_B,
\\
\psi_3=\dfrac{\alpha}{\xi}\ket{\uparrow}_A \otimes \ket{\downarrow}_N\otimes \ket{\uparrow}_B+\dfrac{2a}{\delta}\ket{\downarrow}_A \otimes \ket{\uparrow}_N\otimes \ket{\uparrow}_B,
\\
 \psi_4=\dfrac{-1}{\sqrt{5}}\ket{\uparrow}_A \otimes \ket{\downarrow}_N\otimes \ket{\downarrow}_B-\dfrac{2 \sgn(\alpha)}{\sqrt{5}}\ket{\downarrow}_A \otimes \ket{\uparrow}_N\otimes \ket{\downarrow}_B,
\\
 \psi_5=\dfrac{a}{2\xi}\ket{\uparrow}_A \otimes \ket{\downarrow}_N\otimes \ket{\uparrow}_B-\dfrac{4\alpha}{\delta}\ket{\downarrow}_A \otimes \ket{\uparrow}_N\otimes \ket{\uparrow}_B,
\\
 \psi_6=\dfrac{2}{\sqrt{5}}\ket{\uparrow}_A \otimes \ket{\downarrow}_N\otimes \ket{\downarrow}_B-\dfrac{ \sgn(\eta)}{\sqrt{5}}\ket{\downarrow}_A \otimes \ket{\uparrow}_N\otimes \ket{\downarrow}_B,
\\
\psi_7=\ket{\downarrow}_A \otimes \ket{\downarrow}_N\otimes \ket{\uparrow}_B,
\\
\psi_8=\ket{\downarrow}_A \otimes \ket{\downarrow}_N\otimes \ket{\downarrow}_B.
 \end{gathered}
\end{align}
\end{ceqn}
where $\alpha=B-\frac{a}{4}$, $\xi=\sqrt{|\alpha|^2+ \dfrac{a^2}{4}}$, $\delta=\sqrt{\alpha^2+ 4 a^2}$, and $\eta=4B-3a$.

In this simple model, we assume the radical pairs are initially in a singlet state and the nuclear spin is up. In the rest of this paper and the expression for the singlet yield in this section, the initial nuclear states are averaged over:

\begin{equation}
\dfrac{1}{\sqrt{2}} \Big( \ket{\uparrow}_A \otimes \ket{\downarrow}_B -\ket{\downarrow}_A \otimes \ket{\uparrow}_B\Big)\otimes \ket{\uparrow}_N=\dfrac{1}{\sqrt{2}} \Big( \ket{\uparrow}_A \otimes \ket{\uparrow}_N\otimes \ket{\downarrow}_B -\ket{\downarrow}_A \otimes \ket{\uparrow}_N\otimes \ket{\uparrow}_B\Big).    
\end{equation}

\begin{figure}
 \includegraphics[width=0.55\linewidth]{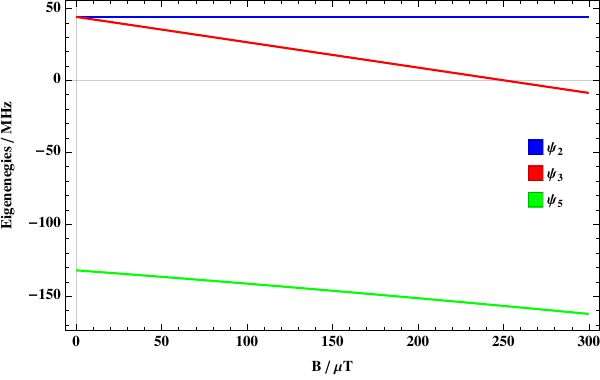}
 \caption{\textbf{Dependence of the eigenenergies on $B$.} Eigenenergies of three states that overlap with the singlet initial state as a function of the applied magnetic field with a hyperfine coupling constant of 1000 $\mu$T.} 
\label{fig:HMF_EG}
\end{figure}

This initial state overlaps with three out of eight eigenstates of the system, $\psi_2$, $\psi_3$, and $\psi_5$. At extremely low fields, two of these three states are degenerate. The third one is split from the other two by the hyperfine interaction, as shown in Fig. \ref{fig:HMF_EG}. Due to this, even in hypomagnetic conditions, this energy difference results in S-T mixing.

\par

Following the work of Timmel et al. \cite{Timmel1998,Hore2019}, the chemical fate of the radical pair can be modelled by separating spin-selective reactions of the singlet and triplet pairs. For simplicity, it is assumed that $k=k_S=k_T$, where $k_S$ and $k_T$ are the singlet and triplet reaction rates, respectively. The final singlet yield, $\Phi^{S(S)}$, for periods much greater than the radical pair lifetime reads as follows:

\begin{ceqn}
\begin{align}
 \Phi^{S(S)} & = k \int_{0}^{\infty} \expval{\hat{P}^S}(t) e^{-kt} dt  ={}\dfrac{1}{4}-\dfrac{k}{4(k+r)}+\dfrac{1}{M} \sum_{m}^{4M} \sum_{n}^{4M} \abs{\bra{m}\hat{P}^S\ket{n}}^2 \dfrac{k(k+r)}{(k+r)^2+(\omega_m -\omega_n)^2},
  \label{eq:SSY}
\end{align}
\end{ceqn}
where the fractional triplet yield can be calculated as $\Phi^{T(S)}=1-\Phi^{S(S)}$. Similarly, if the radical pairs start off in triplet states, the singlet yield fraction reads as follows:\par
\begin{ceqn}
\begin{align}
 \Phi^{S(T)} ={}\dfrac{1}{4}+\dfrac{k}{12(k+r)}-\dfrac{1}{3M} \sum_{m}^{4M} \sum_{n}^{4M} \abs{\bra{m}\hat{P}^S\ket{n}}^2 \dfrac{k(k+r)}{(k+r)^2+(\omega_m -\omega_n)^2}.
  \label{eq:STY}
\end{align}
\end{ceqn}

\begin{figure}
 \includegraphics[width=0.8\linewidth]{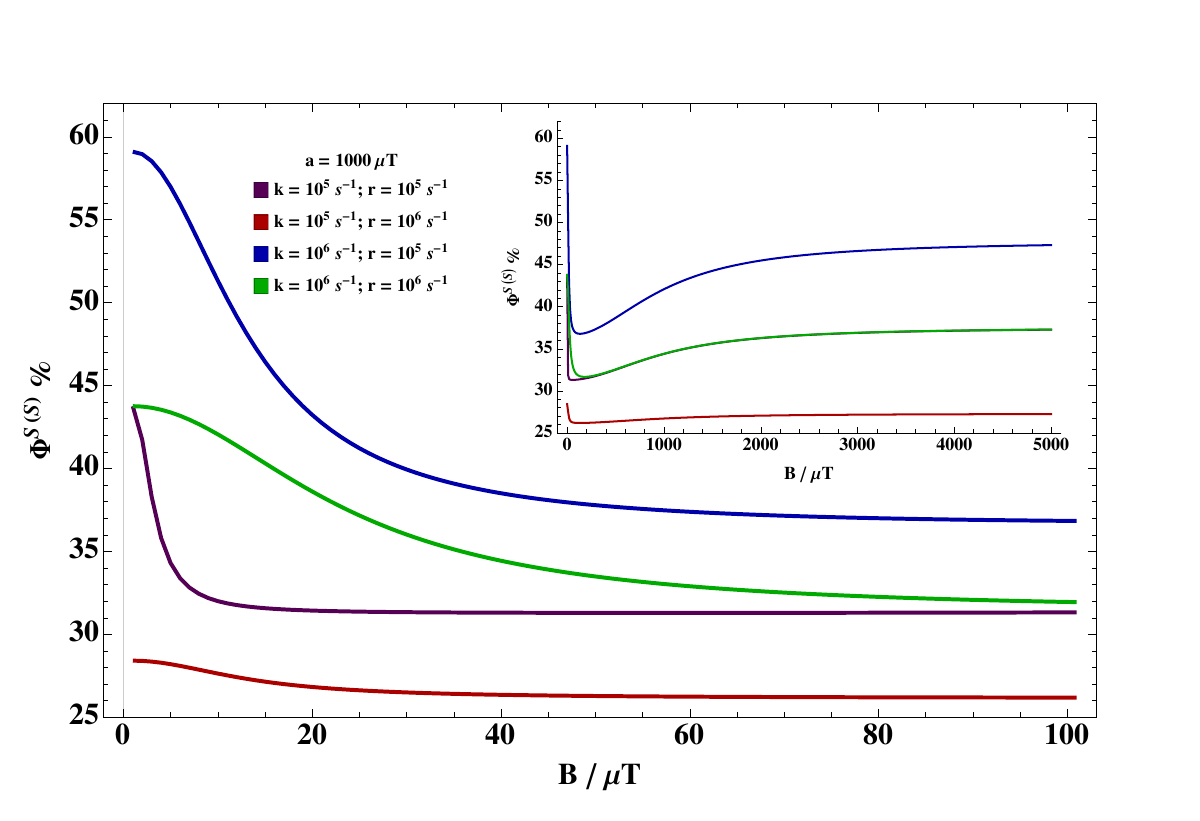}
 \caption{\textbf{Dependence of the fractional singlet yield on $B$.} Dependence of the fractional singlet yield for (a) singlet-born ($\Phi^{S(S)}$) radical pair on the external magnetic field $B$ with one isotropic hyperfine coupling of 1000 $\mu$T for different relaxation rate, $r$, and reaction rates, $k$. There is a peak in the hypomagnetic field range for most values of $k$ and $r$. The peak diminishes when the reaction rate is slower than the spin relaxation rate. One can also see that there is a dip at intermediate field values, then a rise at higher fields, but this rise is not as high as the peak for the hypomagnetic field range (see the inset).} 
\label{fig:SY-B}
\end{figure}

Applied magnetic fields can alter the fractional singlet/triplet yield. Fig. \ref{fig:SY-B} shows the dependence of the fractional singlet yield for singlet-born radical pairs for different sets of relaxation rate, $r$, reaction rate, $k$, and one isotropic hyperfine coupling constant. 
It is clear that there is a peak in the hypomagnetic field range for most values of $k$ and $r$. It is noteworthy that the peak diminishes when the reaction rate is slower than the spin relaxation rate. One can also see that there is a dip at intermediate field values, then a rise at higher fields, but this rise is not as high as the peak for the hypomagnetic field range. In the considered  parameter range, the radical pair mechanism thus predicts quite significant hypomagnetic field effects. It should also be noted that the case of triplet-born and singlet-born radical pairs have the same behavior, but with peaks and dips exchanged.

\subsection{Hypomagnetic field effect}

One can define hypomagnetic field effects, $\Delta \Theta_{HMF\leftrightarrow GMF}$ (similar to previous works on LFEs, e.g., Ref \cite{Timmel1998}), as the absolute value of the ratio of the difference between the singlet (triplet) yield at near zero magnetic field\textcolor{violet}{s} (1 $\mu$T) and at the geomagnetic field range (50 $\mu$T) over the singlet yield at near-zero magnetic field:
\begin{ceqn}
\begin{align}
\Delta\Theta^{i(j)}_{HMF\leftrightarrow GMF}=\abs{\dfrac{\Phi^{i(j)}_{HMF}-\Phi^{i(j)}_{GMF}}{\Phi^{i(j)}_{HMF}}},
\label{eq:HMFE}
\end{align}
\end{ceqn}
where $i,j=\{$S,T$\}$.

We numerically explore the hypomagnetic field effects for the radical pair model using Eq. \ref{eq:HMFE}. This is particularly relevant for experiments based on screening the geomagnetic field (of which there are quite many) \cite{Zadeh_Haghighi_2022}.

\begin{figure}
 \includegraphics[width=0.8\linewidth]{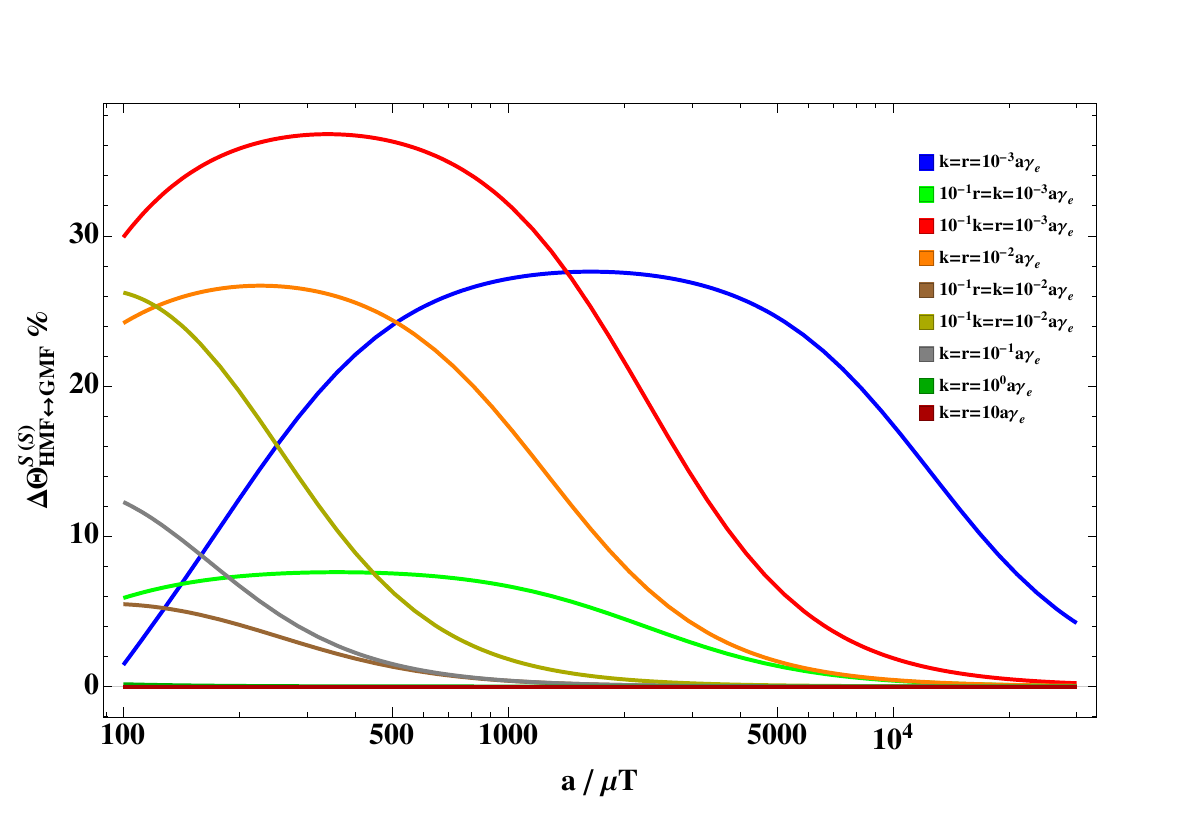}
\caption{\textbf{Dependence of $\Delta \Theta_{HMF\leftrightarrow GMF}^{S(S)}$ on $a$, $r$, $k$.} One of the radicals is coupled to a nucleus with $\dfrac{1}{2}$-spin and the other one with no hyperfine interaction. The hypomagnetic field effects are more evident when the hyperfine coupling is larger than the spin relaxation and reaction rates. This is even more pronounced when the spin relaxation is slower than the reaction rate. If the hyperfine coupling constant is much larger than the reaction and spin relaxation rate, the hypomagnetic field effects can reach considerable values (say $>10\%$) for $a<20000 \mu$T (e.g. Blue curve). $\gamma_e=1.76 \times 10^{11}$ s$^{-1}$ T$^{-1}$ is the electron magnetogyric ratio. Due to the dominance of the hyperfine interaction over the Zeeman interaction, for too high values of the hyperfine coupling constant, there are no hypomagnetic field effects; this depends on the relaxation and reaction rates.}
\label{fig:akr}
\end{figure}

Fig. \ref{fig:akr} shows the dependence of the hypomagnetic field effect on the hyperfine coupling constant for different sets of relaxation and reaction rates. It is worth noting that the hypomagnetic field effects are more evident when the hyperfine coupling is larger than the spin relaxation and reaction rates. This is even more pronounced when the spin relaxation is slower than the reaction rate. It is also worth mentioning that if the hyperfine coupling constant is much larger than the reaction and spin relaxation rate, the hypomagnetic field effects can reach considerable values (say $>10\%$) for $a<20000 \mu$T. For too high values of the hyperfine coupling constant, there are no hypomagnetic field effects. This is due to the fact that, for all different rates, at far too large hyperfine couplings, the hyperfine interaction dominates the Zeeman interaction, and hence there are no quantum beats.     

\begin{figure}
  \includegraphics[width=0.6\linewidth]{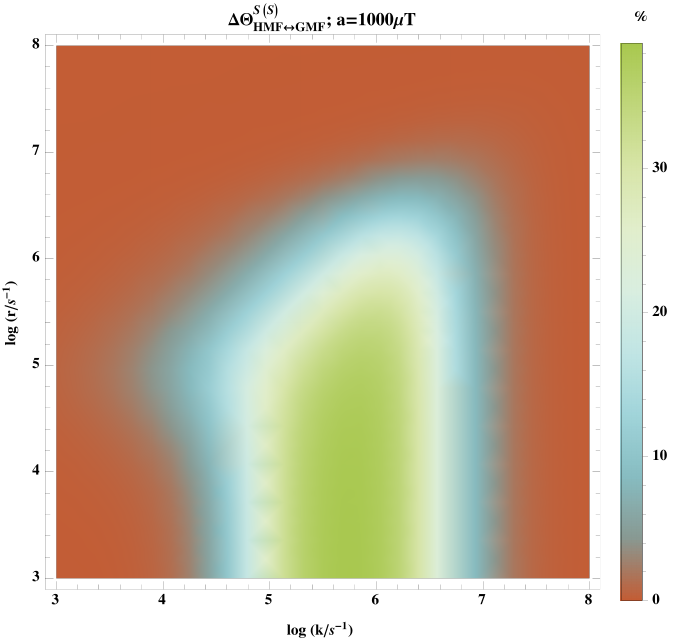}
 \caption{\textbf{Dependence of $\Delta \Theta_{HMF\leftrightarrow GMF}^{S(S)}$ on $k$ and $r$.} The hypomagnetic field effect is based on the singlet yield ratio of 1 $\mu$T (HMF) over 50 $\mu$T (GMF) with singlet initial states. In this model, one of the radicals has no hyperfine interaction, and the other one is coupled to one nucleus with an isotropic hyperfine coupling constant of $a=1000\mu$T. The hypomagnetic field effects are significant when $r<10^6$s$^{-1}$, $k\in[10^{4.5},10^{6.5}]$s$^{-1}$, and $k<r$.}
 \label{fig:HMF_cntrS_1000}
\end{figure}

Fig. \ref{fig:HMF_cntrS_1000} shows the hypomagnetic field effects based on the singlet yield ratio with singlet initial states, $\Phi^{S(S)}$, where that one of the radicals is coupled to one nuclear spin-$\dfrac{1}{2}$ with a hyperfine coupling of 1000 $\mu$T. The effect is higher than 10\% when $k\in [10^5-10^7]$ s$^{-1}$ and $r<10^6$ s$^{-1}$. The maximum value for hypomagnetic field effects for singlet-born radical pairs reaches $\sim$40\%.

\begin{figure}
        \begin{subfigure}[b]{0.4\textwidth}
            \includegraphics[width=1\textwidth]{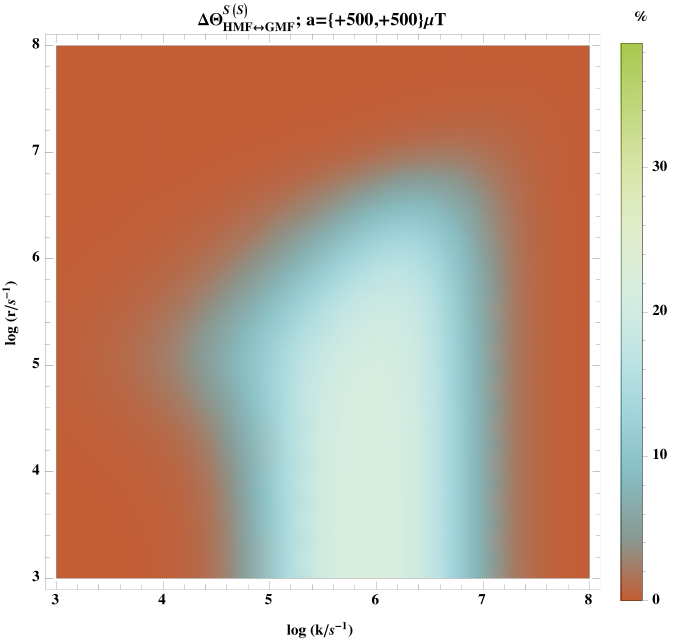}
            \caption{
                \label{fig:StNDdbL}}
        \end{subfigure}
        \hspace{2ex}
        \begin{subfigure}[b]{0.4\textwidth}
            \includegraphics[width=1\textwidth]{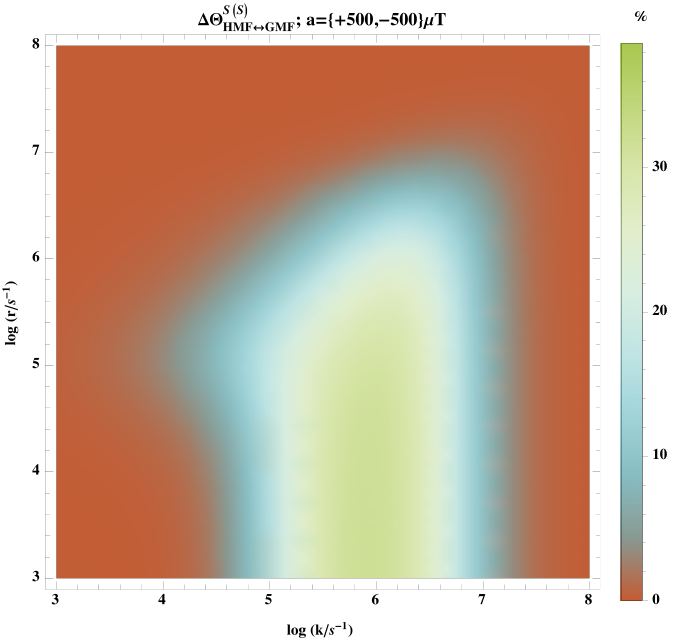}
            \caption{
                \label{fig:StNDdbP}}
        \end{subfigure}
        \\[3ex]
        
        \begin{subfigure}[b]{0.4\textwidth}
            \includegraphics[width=1\textwidth]{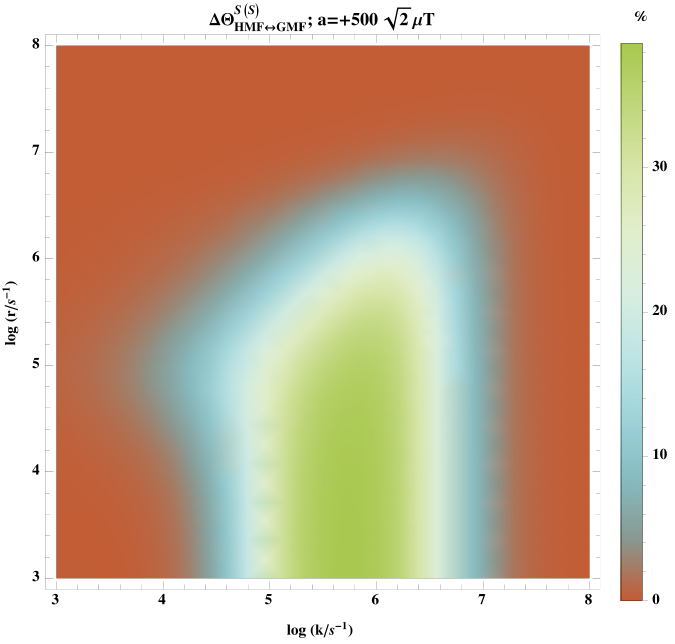}
            \caption{
                \label{fig:StNFbpL}}
        \end{subfigure}
       
        \caption{\textbf{Comparing different signs for the hyperfine couplings in the case of multiple nuclei.} Dependence of $\Delta \Theta_{HMF\leftrightarrow GMF}^{S(S)}$ on $k$, $r$, and signs of the hyperfine coupling constants. The hypomagnetic field effect is based on the singlet yield ratio of 1 $\mu$T (HMF) over $\mu$T (GMF) with initial singlet states. In this model, one of the radicals has no hyperfine interaction and the other one is coupled to two nuclei with these isotropic hyperfine coupling constants: (a) $a_{iso}=\{+500,+500\} \mu$T, (b) $a_{iso}=\{+500,-500\} \mu$T, (c) $a_{eiso}=500\sqrt{2} \mu$T. The signs of the hyperfine couplings have a negligible impact on the fractional yield for the hypomagnetic field effects. The effective coupling constant has a stronger effect than the other coupling constants.}  
\label{fig:HMF_cntrSS}
    \end{figure}

Most organic radicals possess several nuclei carrying spin with positive or negative isotropic hyperfine coupling constant. Fig. \ref{fig:HMF_cntrSS} shows 
the dependence of the hypomagnetic field effect on spin relaxation rate and reaction rate for these sets of isotropic hyperfine coupling constants: $a_{iso}=\{+500, +500\} \mu$T, $a_{iso}=\{+500, -500\} \mu$T, and $a_{eiso}= 500 \sqrt{2} \mu$T. The signs of the hyperfine couplings have a minor impact on the fractional yield for the hypomagnetic field effects. Here, one of the radicals is coupled to two nuclei, and the other radical has no hyperfine interaction. $a_{eiso}$ is the effective hyperfine coupling constant:
\begin{ceqn}
\begin{align}
a_{eiso}=\sqrt{\dfrac{4}{3}\sum_{i}^{N} a_i^2 I_i (I_i+1)}.
\end{align}
\end{ceqn}

The effect is higher than 10\% when $k\in [10^5-10^7]$ s$^{-1}$ and $r<10^6$ s$^{-1}$. In this range of parameters, the effective hyperfine coupling constant shows stronger effects relative to the other sets of couplings.  

In the present work, the second nuclear spin was considered to be on the same radical as the first one, and the other radical had no hyperfine interaction. It should also be noted that when the second nucleus is not in the same radical as 
the first one, the hypomagnetic field effect should be extremely small \cite{Timmel1998}.
\par
\begin{figure}
        \begin{subfigure}[b]{0.4\textwidth}
            \includegraphics[width=1\textwidth]{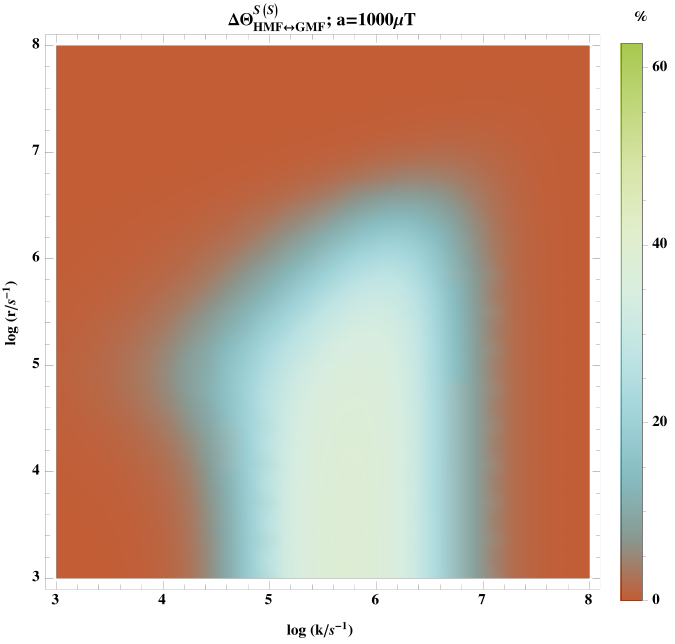}
            \caption{
                \label{fig:StNDdbLs}}
        \end{subfigure}
        \hspace{2ex}
        \begin{subfigure}[b]{0.4\textwidth}
            \includegraphics[width=1\textwidth]{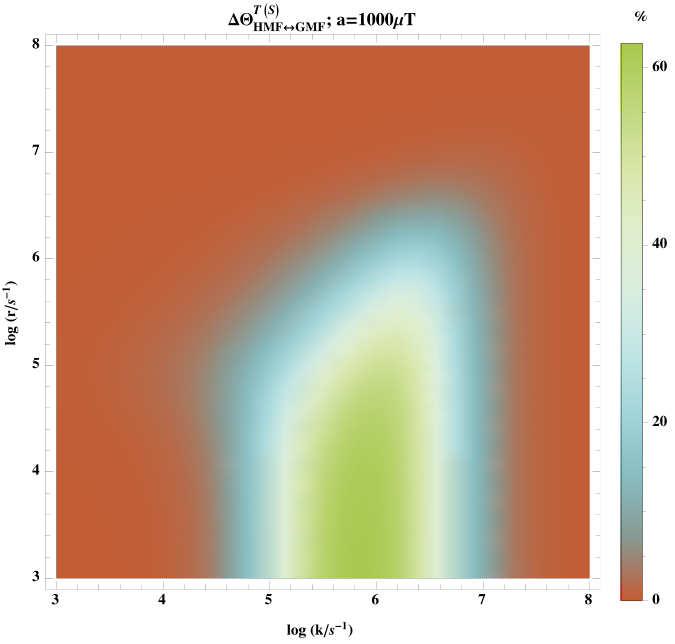}
            \caption{
                \label{fig:StNDdbPs}}
        \end{subfigure}
        \\[3ex]
        
        \begin{subfigure}[b]{0.4\textwidth}
            \includegraphics[width=1\textwidth]{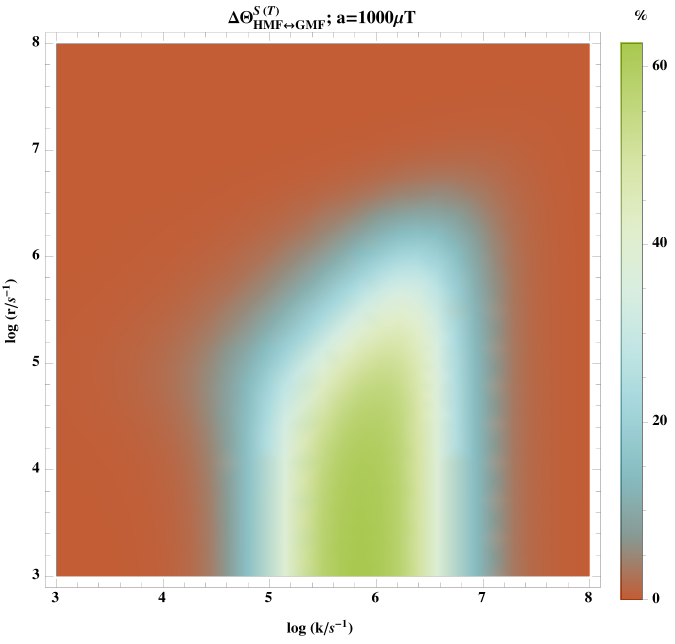}
            \caption{
                \label{fig:StNFbpLs}}
        \end{subfigure}
        \hspace{2ex}
        \begin{subfigure}[b]{0.4\textwidth}
            \includegraphics[width=1\textwidth]{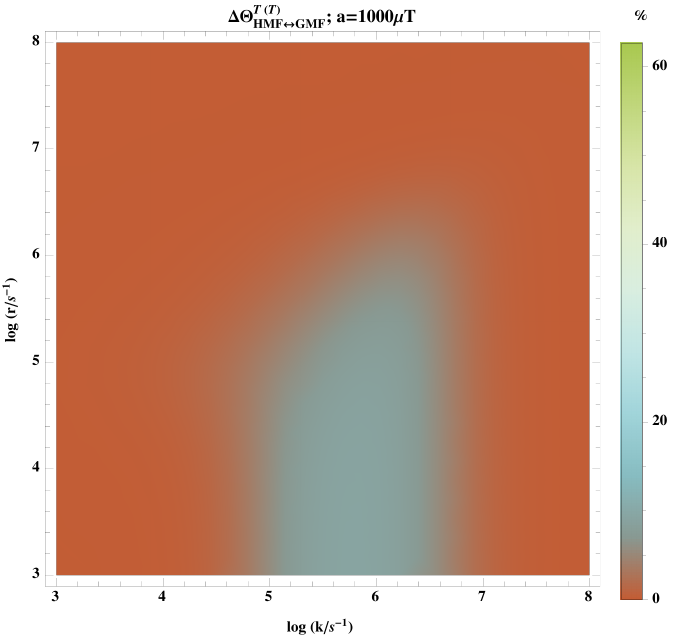}
            \caption{
                \label{fig:StNFbpPx}}
        \end{subfigure}
        \caption{\textbf{Impact of the initial state of the radical pairs (singlet or triplet) on the hypomagnetic field effects.} The hypomagnetic field effect is based on the singlet yield ratio with initial singlet states (a), the triplet yield ratio with initial singlet states (b), the singlet yield ratio with initial triplet states (c), and the triplet yield ratio with initial triplet states (d). In this model, one of the radicals has no hyperfine interaction, and the other one is coupled to an isotropic hyperfine coupling constant of $a_{eiso}=1000 \mu$T. The effect is evident when $k\in [10^5-10^7]$ s$^{-1}$ and $r<10^6$ s$^{-1}$. However, the effect is more pronounced when is based on singlet (triplet) yield with triplet (singlet) initial states, $\Phi^{S(T)}$($\Phi^{T(S)}$).
            \label{fig:initialstate}}
    \end{figure}

We also investigate how the initial state of the radical pairs (singlet or triplet) impacts the hypomagnetic field effects. Fig. \ref{fig:initialstate} shows the hypomagnetic field effects based on singlet/triplet yield with singlet/triplet initial states. The effect is evident when $k\in [10^5-10^7]$ s$^{-1}$ and $r<10^6$ s$^{-1}$. However, the effect is more pronounced when it is based on singlet (triplet) yield with triplet (singlet) initial states, $\Phi^{S(T)}$($\Phi^{T(S)}$),  whereas the effect based on triplet yield with triplet initial states, $\Phi^{T(T)}$, is less than 10\%. 

\begin{figure}
        \begin{subfigure}[b]{0.4\textwidth}
            \includegraphics[width=1\textwidth]{Figure4.pdf}
            \caption{
                \label{fig:StNDcdbLs}}
        \end{subfigure}
        \hspace{2ex}
        \begin{subfigure}[b]{0.4\textwidth}
            \includegraphics[width=1\textwidth]{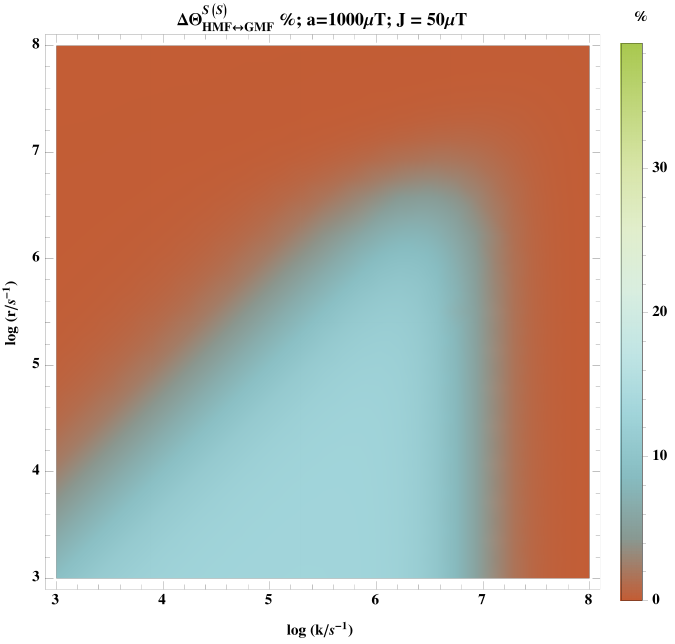}
            \caption{
                \label{fig:StNDdcbPs}}
        \end{subfigure}
        \caption{\textbf{Effect of exchange interaction on the hypomagnetic field effects.} The hypomagnetic field effect is based on the singlet yield ratio with initial singlet states without (a) and with (b) the exchange interaction. The exchange interaction coupling is -50$\mu$T. In this model, one of the radicals has no hyperfine interaction, and the other is coupled to an isotropic hyperfine coupling constant of $a_{eiso}=1000 \mu$T. The exchange interaction significantly attenuates the hypomagnetic field effects.
            \label{fig:exchange}}
    \end{figure}

We also considered the effects of the exchange interaction: 
\begin{ceqn}
\begin{align}
H_J=J \mathbf{\hat{S}}_A\sbullet\mathbf{\hat{S}}_B
\end{align}
\end{ceqn}
where $J$ is the exchange coupling. Fig. \ref{fig:exchange} shows that coupling of -50 $\mu$T can significantly attenuate the hypomagnetic field effects. A priori for low field effects inter-radical spin-spin interactions would need to be $\leq$50 $\mu$T, which requires $\geq$3.5 nm distance between the radicals. This challenges the radical pair mechanism because electron transfer may be slower than spin relaxation for relatively large distances.  However, Ref \cite{Efimova2008} showed that LFEs can be significant for much smaller separation of order 2 nm where the effects of exchange and dipolar interactions partially cancel.

We also explored the effects of small fluctuations on the hypomagnetic field. We considered an oscillating magnetic field with field with a low frequency and intensity: 
\begin{ceqn}
\begin{equation}
B=B_0+B_1 \cos{\alpha},
\label{eq:osc1}
\end{equation}
\end{ceqn}
where $B_1$ is the small fluctuating field. As the lifetime of radical pairs is much shorter than  the frequency of the applied magnetic field, the effect of the fluctuation can be treated as static during the lifetime of a radical pair \cite{Scaiano1994,Hore2019}. 

\begin{ceqn}
\begin{equation}
    \overline{\Phi_S(B_0,B_1)}=\frac{1}{\pi}\int_0^\pi\Phi_S(B) d\alpha.
    \label{eq:osc2}
\end{equation}
\end{ceqn}

\begin{figure}
        \begin{subfigure}[b]{0.4\textwidth}
            \includegraphics[width=1\textwidth]{Figure4.pdf}
            \caption{
                \label{fig:StNfgfgDcdbLs}}
        \end{subfigure}
        \hspace{2ex}
        \begin{subfigure}[b]{0.4\textwidth}
            \includegraphics[width=1\textwidth]{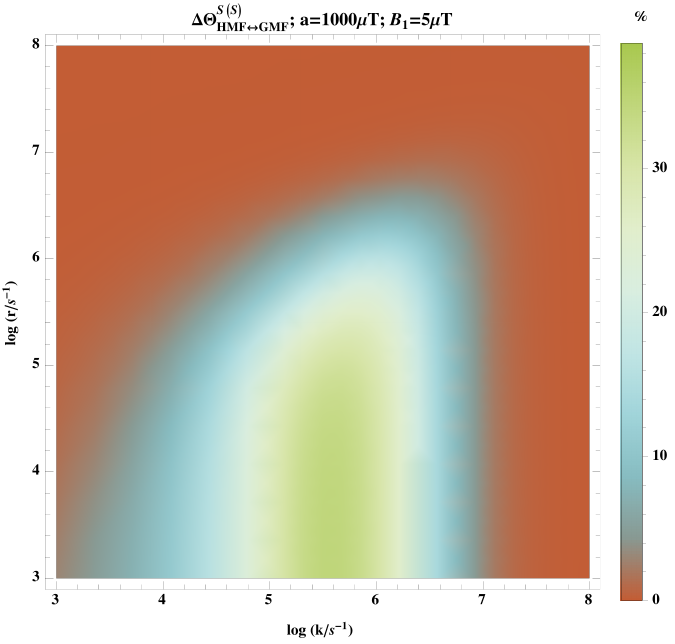}
            \caption{
                \label{fig:StNDddfdfcbPs}}
        \end{subfigure}
        \caption{\textbf{Effect of small fluctuation on the hypomagnetic field effects.} The hypomagnetic field effect is based on the singlet yield ratio with initial singlet states without (a) and with (b) fluctuating magnetic field of 5 $\mu$T with low frequency. In this model, one of the radicals has no hyperfine interaction, and the other one is coupled to an isotropic hyperfine coupling constant of $a_{eiso}=1000 \mu$T. The fluctuating field does not change the hypomagnetic field effects significantly.
            \label{fig:fluc}}
    \end{figure}

Fig. \ref{fig:fluc} shows the effect of 5 $\mu$T fluctuating magnetic field with low frequency on the hypomagnetic field effect, where including this does not change the results significantly.

\section{Discussion}

In the present work, we aimed to show the hypomagnetic field effects may be an interesting platform for testing the radical pairs models as the underlying mechanism behind the weak magnetic fields in biology. The results indicate that the response of biological systems going from the geomagnetic field strength to the hypomagnetic field regime may be more pronounced than going from the geomagnetic field to higher fields, as shown in Fig. \ref{fig:SY-B}, which shows clearly that the peak at the low field can be higher than the effect of strong fields. The peak diminishes when the reaction rate is slower than the spin relaxation rate. The hypomagnetic field effects are also more evident when the hyperfine coupling is larger than the spin relaxation and reaction rates. The effects exhibit negligible dependence on the signs of the hyperfine coupling constants, where the effective coupling constant has a larger effective size relative to the individual hyperfine interactions for multi-nuclei interaction. We also explored the impact of the initial state of the radical pairs (singlet or triplet) on the hypomagnetic field effects. We concluded that the effect is more pronounced when it is based on singlet (triplet) yield with triplet (singlet) initial states. We also showed that the exchange interaction can significantly attenuate the hypomagnetic field effects. The effect of small fluctuating magnetic fields on the hypomagnetic field effect was insignificant.  

It should be noted that the radical pair model considered in the present work is simplified, where the reaction rates for singlet and triplet yields and the spin relaxation for both radicals are assumed equal. More realistic models of the radical pairs may provide further insight into the underlying mechanism behind these phenomena, e.g., solving the master equation and considering the role of entanglement, dipolar, and exchange interactions in the models. 
 
It should also be noted that an extended version of the radical pair mechanism, the radical triad mechanism \cite{Kattnig2017,Kattnig2017a,Ramsay_2022}, has been proposed as an explanation for weak magnetic field effects. Radical triads may provide more sensitivity and circumvent issues such as weakening effects from dipolar interactions and fast spin relaxation of superoxide. There are several pieces of evidence that radical pairs or triads may be involved in the response of biological systems to weak magnetic fields \cite{Xu2021,Hiscock2016}. However, both approaches demand more supporting evidence, and this question is under active investigation.

A considerable amount of evidence shows that shielding the geomagnetic field has direct biological consequences, which in some cases could be detrimental. This could also be pertinent for the quest for life on other planets with different magnetic fields, including Mars, which has zero magnetic fields \cite{McKay1996,Hyodo2021}.

Studies suggest the Laschamp excursion--the most intense geomagnetic event occurring over the last 50 kyr, with a quasi-reversed polarity of the geomagnetic field--in combination with the Grand Solar Minima, initiated substantial changes in the concentration and circulation of the atmospheric ozone, increased atmospheric ionisation and ultraviolet radiation levels, leading to global climate shifts that caused major environmental changes \cite{Cooper2021}. They concluded the excursions of the geomagnetic field could be potentially harmful to life. In other words, they proposed an indirect impact of the geomagnetic field deprivation on the mass extinction indirectly via the atmosphere. 

In the current work, we reviewed the radical pair mechanism as an arguably plausible explanation for hypomagnetic field effects in biology. We hope this work will inspire our experimental colleagues to test the credibility of this explanation.

\section*{Data Availability}
The generated datasets and computational analysis are available from the corresponding author upon reasonable request.

\section*{Acknowledgment}

The authors would like to thank Aurélien de la Lande, Jean Deviers, and Dennis Salahub for their valuable input. This work was supported by the Natural Sciences and Engineering Research Council of Canada.

\section*{Author Contributions}

H.Z. and C.S. conceived the project. H.Z. performed formal analysis, investigation, methodology, software, and visualization with feedback from C.S. H.Z. wrote the original draft with help from C.S. and R.R.

\bibliography{sample}

\end{document}